\documentclass[reprint,superscriptaddress,aps,prl]{revtex4-1}

\usepackage{graphicx}
\usepackage{dcolumn}
\usepackage{bm}
\usepackage{bbold}
\usepackage{amsfonts,color}
\usepackage{amsmath}

\newcounter{lastnote}

\begin{document}

\title{Chiral spin-wave velocities induced by all-garnet interfacial Dzyaloshinskii-Moriya interaction in ultrathin yttrium iron garnet films}

\author{Hanchen Wang}
\thanks{These authors contributed equally to this work.}
\affiliation{%
Fert Beijing Institute, BDBC, School of Microelectronics, Beihang University, Beijing 100191, China
}
\author{Jilei Chen}
\thanks{These authors contributed equally to this work.}
\affiliation{%
Fert Beijing Institute, BDBC, School of Microelectronics, Beihang University, Beijing 100191, China
}
\affiliation{%
Laboratory of Nanoscale Magnetic Materials and Magnonics,~Institute of Materials (IMX),~School of Engineering, \'Ecole Polytechnique F\'ed\'erale de Lausanne (EPFL), 1015 Lausanne, Switzerland
}%
\author{Tao Liu}
\thanks{These authors contributed equally to this work.}
\affiliation{%
Department of Physics, Colorado State University, Fort Collins, Colorado 80523, USA
}%
\author{Jianyu Zhang}
\affiliation{%
Fert Beijing Institute, BDBC, School of Microelectronics, Beihang University, Beijing 100191, China
}
\author{Korbinian~Baumgaertl}
\affiliation{%
Laboratory of Nanoscale Magnetic Materials and Magnonics,~Institute of Materials (IMX),~School of Engineering, \'Ecole Polytechnique F\'ed\'erale de Lausanne (EPFL), 1015 Lausanne, Switzerland
}%
\author{Chenyang Guo}
\affiliation{%
Beijing National Laboratory for Condensed Matter Physics, Institute of Physics, University of Chinese Academy of Sciences, Chinese Academy of Sciences, Beijing 100190, China
}%
\author{Yuehui Li}
\affiliation{%
Electron Microscopy Laboratory, School of Physics, Peking University, Beijing 100871, China
}%
\affiliation{%
International Center for Quantum Materials, School of Physics, Peking University, Beijing 100871, China
}%
\author{Chuanpu Liu}
\affiliation{%
Fert Beijing Institute, BDBC, School of Microelectronics, Beihang University, Beijing 100191, China
}
\affiliation{%
Department of Physics, Colorado State University, Fort Collins, Colorado 80523, USA
}
\author{Ping Che}
\affiliation{%
Laboratory of Nanoscale Magnetic Materials and Magnonics,~Institute of Materials (IMX),~School of Engineering, \'Ecole Polytechnique F\'ed\'erale de Lausanne (EPFL), 1015 Lausanne, Switzerland
}
\author{Sa Tu}
\affiliation{%
Fert Beijing Institute, BDBC, School of Microelectronics, Beihang University, Beijing 100191, China
}%
\author{Song Liu}
\affiliation{%
Shenzhen Institute for Quantum Science and Engineering (SIQSE), and Department of Physics, Southern University of Science and Technology (SUSTech), Shenzhen 518055, China
}%
\author{Peng Gao}
\affiliation{%
Electron Microscopy Laboratory, School of Physics, Peking University, Beijing 100871, China
}%
\affiliation{%
International Center for Quantum Materials, School of Physics, Peking University, Beijing 100871, China
}%
\affiliation{%
Collaborative Innovation Center of Quantum Matter, Beijing 100871, China
}%
\author{Xiufeng~Han}
\affiliation{%
Beijing National Laboratory for Condensed Matter Physics, Institute of Physics, University of Chinese Academy of Sciences, Chinese Academy of Sciences, Beijing 100190, China
}%
\author{Dapeng~Yu}
\affiliation{%
Shenzhen Institute for Quantum Science and Engineering (SIQSE), and Department of Physics, Southern University of Science and Technology (SUSTech), Shenzhen 518055, China
}%
\affiliation{%
Electron Microscopy Laboratory, School of Physics, Peking University, Beijing 100871, China
}%
\author{Mingzhong~Wu}%
\affiliation{%
Department of Physics, Colorado State University, Fort Collins, Colorado 80523, USA
}%
\author{Dirk Grundler}%
\affiliation{%
Laboratory of Nanoscale Magnetic Materials and Magnonics,~Institute of Materials (IMX),~School of Engineering, \'Ecole Polytechnique F\'ed\'erale de Lausanne (EPFL), 1015 Lausanne, Switzerland
}%
\affiliation{%
Institute of Microengineering (IMT), School of Engineering, \'Ecole Polytechnique F\'ed\'erale de Lausanne (EPFL), 1015 Lausanne, Switzerland
}%
\author{Haiming Yu}
\email{haiming.yu@buaa.edu.cn}
\affiliation{%
Fert Beijing Institute, BDBC, School of Microelectronics, Beihang University, Beijing 100191, China
}
\date{\today}

\begin{abstract}
Spin waves can probe the Dzyaloshinskii-Moriya interaction (DMI) which gives rise to topological spin textures, such as skyrmions. However, the DMI has not yet been reported in yttrium iron garnet (YIG) with arguably the lowest damping for spin waves. In this work, we experimentally evidence the interfacial DMI in a 7~nm-thick YIG film by measuring the nonreciprocal spin-wave propagation in terms of frequency, amplitude and most importantly group velocities using all electrical spin-wave spectroscopy. The velocities of propagating spin waves show chirality among three vectors, i.e. the film normal direction, applied field and spin-wave wavevector. By measuring the asymmetric group velocities, we extract a DMI constant of 16~$\mu$J/m$^{2}$ which we independently confirm by Brillouin light scattering. Thickness-dependent measurements reveal that the DMI originates from the oxide interface between the YIG and garnet substrate. The interfacial DMI discovered in the ultrathin YIG films is of key importance for functional chiral magnonics as ultra-low spin-wave damping can be achieved.
\end{abstract}


\maketitle

\begin{figure}[!ht]
\centering
\includegraphics[width=86mm]{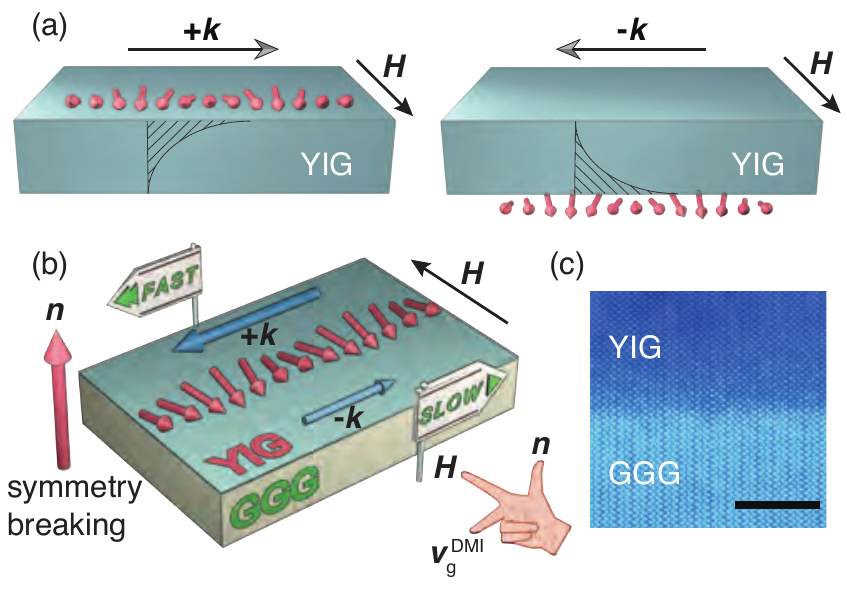}
\caption{(a) Damon-Eshbach spin-wave chirality in thick YIG films. (b) An illustrative diagram of the chiral propagation of spin waves in an ultrathin YIG film. The group velocities of spin waves propagating in $+k$ and $-k$ directions are different. The inset shows a right-handed chirality among three vectors, i.e. the film normal direction $\emph{n}$, applied field $\emph{H}$ and DMI-induced drift group velocity $v_{\text{g}}^{\text{DMI}}$. (c) A high-angle annular dark-field image is taken at the YIG/GGG interface of the 7 nm-thick YIG sample. The scale bar is 5 nm.}
\label{fig1}
\end{figure}

\begin{figure}[!ht]
\centering
\includegraphics[width=75mm]{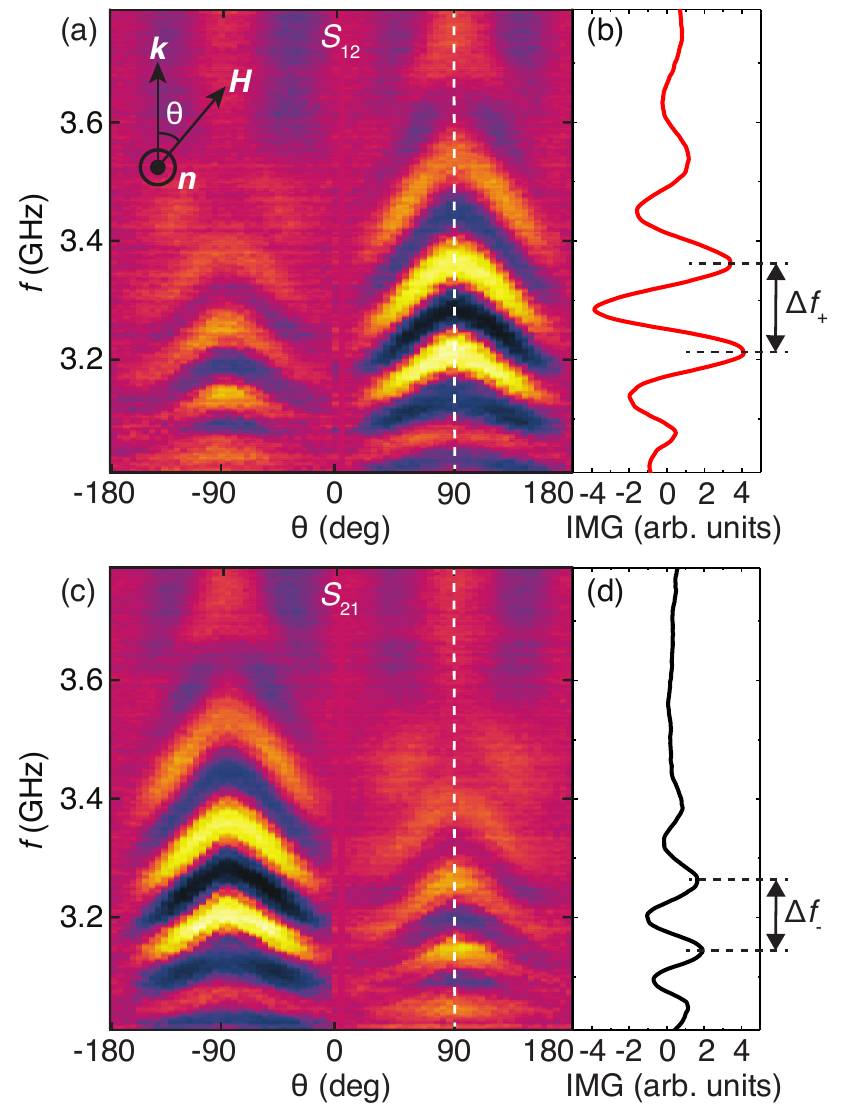}
\caption{Spin-wave transmission spectra $S_{12}$ (a) and $S_{21}$ (c) measured as a function of the field angle $\theta$. The field is fixed at 44 mT. $\theta$ is defined as the angle between $H$ and $k$. Single spectra at 90$^\circ$ (dashed lines in (a) and (c)) are shown for $S_{12}$ (b) and $S_{21}$ (d). The peak-to-peak frequencies $\Delta f_{+}\approx 0.15$ GHz and  $\Delta f_{-}\approx 0.11$ GHz are extracted for the estimation of spin-wave group velocities $v_{\text{g}}^{+}$ and $v_{\text{g}}^{-}$. The 260 nm-wide striplines are used in the experiments~\cite{SI}.}
\label{fig2}
\end{figure}

\begin{figure}[!ht]
\centering
\includegraphics[width=72mm]{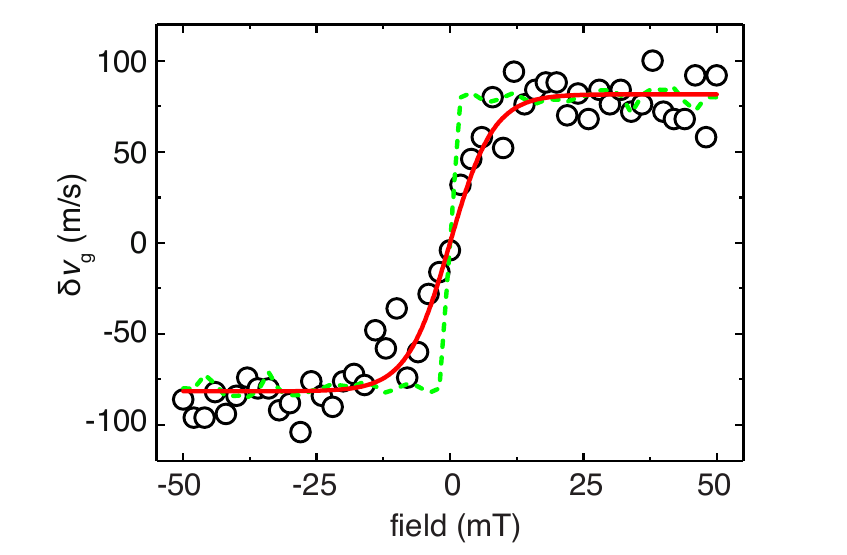}
\caption{The asymmetric group velocity $\delta v_{\text{g}}=v_{\text{g}}^{+}-v_{\text{g}}^{-}$ as a function of the applied field. The field is swept from -50 mT to 50 mT. Black circles are data points calculated using the values of $v_{\text{g}}^{+}$ and $v_{\text{g}}^{-}$ extracted from experiments~\cite{SI}. The red line is a fit using Eq.~\ref{fitting}. The green dashed line is the micromagnetic simulation results~\cite{SI}.}
\label{fig3}
\end{figure}

\begin{figure}[!ht]
\includegraphics[width=86mm]{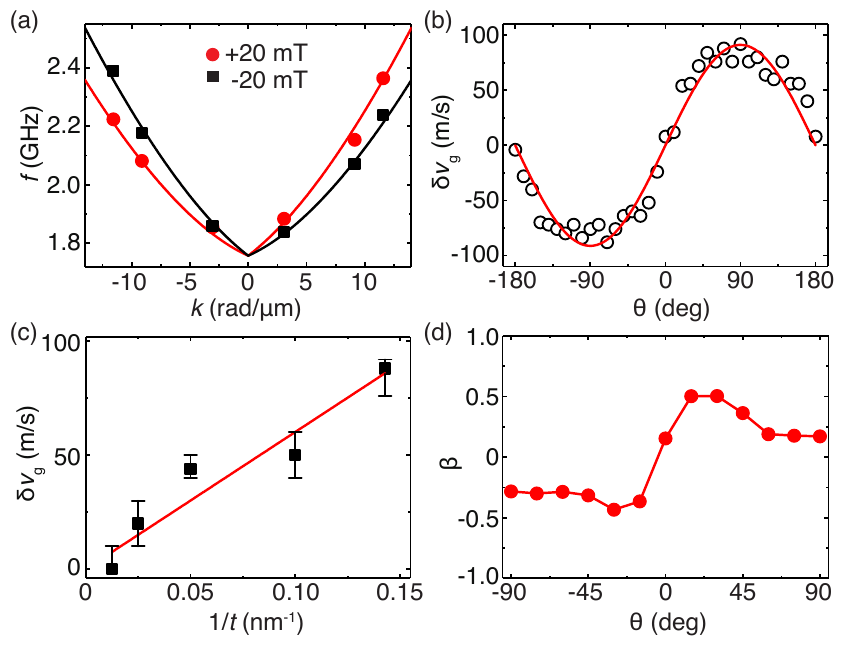}
\centering
\caption{(a) The asymmetric spin-wave dispersion in the presence of an interfacial DMI. The red dots (black squares) are experimental data extracted with an applied field of +20 mT (-20 mT). The red line (black line) is the calculated dispersion using a DMI constant of 16~$\mu$J/m$^2$ for +20 mT (-20 mT). (b) Angle-dependent asymmetric group velocities $\delta v_{\text{g}}$. $\theta$ is defined in Fig.~\ref{fig2}(a). The field is set at 44~mT. Black open circles are data points extracted from the experiments and the red line is a sinusoidal fitting. (c) $\delta v_{\text{g}}$ for samples with different thicknesses of 7 nm, 10 nm, 20 nm, 40 nm and 80 nm with an applied field of 20 mT. The red line is a linear fit to the experimental data. (d) Spin-wave amplitude nonreciprocity $\beta$ measured as a function of $\theta$ on the 10 nm-thick sample. The field is fixed at 10~mT. The 730 nm-wide striplines are used in the experiments~\cite{SI}.}
\label{fig4}
\end{figure}

\begin{figure}[!ht]
\includegraphics[width=70mm]{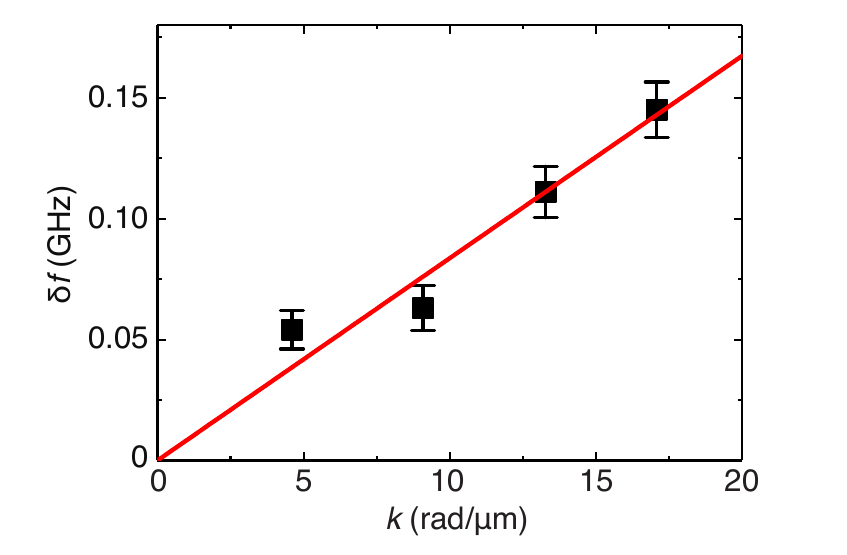}
\caption{Frequency shift $\delta f$ (black squares) measured on the 10 nm-thick YIG sample in an applied field of 80 mT with BLS in reflection geometry at four different incident angles probing spin waves with wavevectors $k=4.6$ rad/$\mu$m, 9.1 rad/$\mu$m, 13.3 rad/$\mu$m and 17.1 rad/$\mu$m~\cite{SI}. The red line is a linear fit to the experimental data.}
\label{fig5}
\end{figure}

Spin waves (or magnons)~\cite{Kru2010,Chu2015,Dem2017} are collective magnetic excitations that can propagate in metals~\cite{Vla2008} and also in insulators~\cite{Sai2010}. Intensive research efforts have been made to investigate spin waves in yttrium iron garnet Y$_3$Fe$_5$O$_{12}$ (YIG)~\cite{Ser2010,vanWees2015} which exhibits the lowest damping that is promising for low-power consumption magnonic devices~\cite{Schn2008,Khitun2010,Csa2016}. Most previous works were conducted on bulk or thick YIG films where the Damon-Eshbach (DE) spin-wave chirality~\cite{DE1961,Demi2009,Saitoh2019} is well known for magnetostatic surface spin waves as illustrated in Fig.~\ref{fig1}(a). However, the DE spin-wave chirality was negligible~\cite{Wong2014} in the thin YIG films which were recently achieved with high quality for on-chip magnonic devices~\cite{Chang2014,Yu2014}. We report a different type of spin-wave chirality scaled up in ultrathin YIG films, which we attribute to the interfacial Dzyaloshinskii-Moriya interaction (DMI)~\cite{Dzy1958,Mor1960}. Very recently, domain wall motion in Tm$_3$Fe$_5$O$_{12}$ (TmIG) on gadolinium gallium garnet (GGG) suggested interfacial DMI consistent with the Rashba effect at oxide-oxide interfaces \cite{Beach2019,Ding2019,Velez2019}. However, independent evidence for the DMI was not provided. Spin waves provide a prevailing methodology~\cite{Zak2010,Cor2013,Moon2013,Nem2015,Lee2016,Li2017,Luc2019} for probing the DMI in metallic multilayers. DMI is the fundamental mechanism to form chiral spin textures, such as skyrmions~\cite{Muh2009,Tokura2010,Sam2013}. Most previous studies on the interfacial DMI focus on measuring frequency shifts $\delta f$ between counter-propagating spin waves (with wavevectors $+k$ and $-k$) using Brillouin light scattering (BLS)~\cite{Nem2015,Li2017} or all-electrical spin-wave spectroscopy (AESWS)~\cite{Lee2016,Luc2019}. It has been theoretically predicted that the interfacial DMI can generate not only a frequency shift, but also asymmetric spin-wave group velocities~\cite{Moon2013}. So far, there has been no experimental observation of nonreciprocal spin-wave characteristics and group velocities induced by interfacial DMI in bare ultrathin YIG films on GGG.
\\
\indent In this letter, we report chiral propagation of spin waves in a 7 nm-thick YIG film on a (111) GGG substrate~\cite{Liu2014}. The spin waves propagating in the chirally-favored direction are found to be substantially faster than in its counter direction as illustrated in Fig.~\ref{fig1}(b). The asymmetry in group velocities $\delta v_{\text{g}}$ is characterized by AESWS~\cite{Vla2008,Yu2014,Lee2016,Neu2010} to be approximately 80 m/s. The chiral spin-wave velocities in YIG films can be accounted for by an interfacial DMI and a DMI constant of 16~$\mu$J/m$^2$ estimated from the experiments. By integrating different antennas, we vary the spin-wave wavevectors and obtain an asymmetric spin-wave dispersion that can be fitted well using the DMI constant extracted from the chiral spin-wave velocities. Five thin-film YIG samples with thicknesses of 7 nm, 10 nm, 20 nm, 40 nm and 80 nm are investigated. The DMI-induced $\delta f$~\cite{Cor2013,Moon2013} and $\delta v_{\text{g}}$ increase when the film thickness decreases, which demonstrates that the DMI in YIG is of interfacial type. We evidence the DMI in the ultrathin YIG films independently by BLS revealing nonreciprocal spin-wave dispersion relations. The DMI constants extracted from the BLS and the AESWS measurements performed on the 10 nm-thick YIG sample agree well. Our discovery makes chiral magnonics~\cite{Kim2016,Hrabec2017,Mulkers2018} a realistic vision as bare YIG offers ultra-low spin-wave damping and thereby enables the plethora of suggested devices that functionalize for instance unidirectional power flow, magnon Hall effect and nontrivial refraction of spin waves.\\
\indent The YIG films were grown on GGG substrates by magnetron sputtering~\cite{Liu2014}. The damping parameter for the 7 nm-thick YIG film is extracted from the ferromagnetic resonance measurements to be $\alpha=6$.4$\pm2.1\times 10^{-4}$~\cite{SI}. The DE spin-wave chirality~\cite{DE1961,Demi2009,Saitoh2019} [Fig.~\ref{fig1}(a)] is negligible when the thickness is 7 nm~\cite{Wong2014}. However, a new type of spin-wave chirality might arise in the presence of DMI where spin waves propagating in opposite directions not only show amplitude nonreciprocity and frequency shifts, but also chiral spin-wave velocities. They are attributed to a DMI-induced drift group velocity whose direction follows a right-handed rule [inset of Fig.~\ref{fig1}(b)]. Figure~\ref{fig1}(c) shows a high-angle annular dark-field image near the YIG/GGG interface. To measure the spin wave group velocities, two nano-stripelines (NSLs) are integrated on the 7 nm-thick YIG film to excite and detect spin waves~\cite{Ciu2016,SI}. The spin-wave propagation distance $s=2~\mu$m. The spin-wave transmission spectra $S_{12}$ ($S_{21}$) suggest that spin waves are excited by NSL2 (NSL1) and detected by NSL1 (NSL2), which is defined as $+k$ ($-k$) spin-wave propagating directions~\cite{SI}. The measured spectra with an external field swept from -50 mT to 50 mT are shown in Supplementary Material Fig. S3~\cite{SI}. Chiral propagation of spin waves is clearly observed with respect to the wavevector \emph{k} and applied field \emph{H}. This chirality is manifested in the angle-dependent measurement shown in Fig.~\ref{fig2}. Figure~\ref{fig2}(a) shows angle-dependent spectra for $S_{12}$ ($+k$), where clear asymmetry is observed. The transmission spectra show contrast oscillations that indicate the phase variation of propagating spin waves~\cite{Vla2008,Neu2010,Yu2014}. In Fig.~\ref{fig2}(b), we show a single spectrum at 90$^\circ$, where a peak-to-peak frequency span $\Delta f_{\text{+}}$ is marked indicating a phase change of 2$\pi$. According to
\begin{equation}\label{eq1}
v_{\text{g}}^{+(-)}=\frac{d\omega}{dk}=\frac{2\pi\Delta f_{+(-)}}{2\pi/s}=\Delta f_{+(-)}\,s,
\end{equation}
the group velocity $v_{\text{g}}^{+}$ for $+k$ spin waves is estimated as $v_{\text{g}}^{+}\approx 312$~m/s, and $v_{\text{g}}^{-}$ for $-k$ spin waves is estimated as $v_{\text{g}}^{-}\approx 236$~m/s. Interestingly, for -90$^\circ$, i.e. a reversed applied field $H$, the situation is nearly mirrored (Fig.~\ref{fig2}). This demonstrates that spin waves propagating in two opposite directions ($+k$ and $-k$) exhibit different group velocities which can be reversed by changing the sign of the applied field and thereby chiral spin-wave velocities are observed.
\\
\indent Figure~\ref{fig3} shows $\delta v_{\text{g}}=v_{\text{g}}^{+}-v_{\text{g}}^{-}$ extracted from the field-dependent measurements~\cite{SI} based on Eq.~\ref{eq1} as a function of the field applied in 90$^\circ$. Near zero field, the group velocities are reciprocal. However, at positive fields $S_{12}$ is faster than $S_{21}$ and at negative fields $S_{21}$ is faster than $S_{12}$, i.e. chiral spin-wave velocities are observed~\cite{SI}. The $\delta v_{\text{g}}$ shows a distinctive step-like field dependence. Theoretical studies~\cite{Moon2013} have predicted that interfacial DMI can introduce a drift group velocity $v_{\text{g}}^{\text{DMI}}$. We indeed observe $v_{\text{g}}^{\text{DMI}}$ experimentally and extract $v_{\text{g}}^{\text{DMI}}=40.8$ m/s by fitting the experimental results in Fig.~\ref{fig3} with an empirical equation
\begin{equation}\label{fitting}
\delta v_{\text{g}}=2v_{\text{g}}^{\text{DMI}}\text{tanh}\left(\frac{H}{H_{0}}\right),
\end{equation}
where $H_{0}$ is a fitting field value above which the velocity difference saturates. The sign of $v_{\text{g}}^{\text{DMI}}$ is determined by a chiral relation among three unit vectors, i.e. the film normal vector $\hat{\textbf{n}}$, applied field $\hat{\textbf{H}}$, and spin-wave wavevector $\hat{\textbf{k}}$ [inset of Fig.~\ref{fig1}(b)]. According to previous theoretical studies~\cite{Moon2013,Kim2016}, one can write
\begin{equation}\label{chiral}
v_{\text{g}}^{\text{DMI}}=\left[\left(\hat{\textbf{n}}\times\hat{\textbf{H}}\right)\cdot\hat{\textbf{k}}\right]\frac{2\gamma}{M_{\text{S}}}D,
\end{equation}
where $D$ the DMI constant, $M_{\text{S}}$ saturation magnetization. At a positive field, for example, the drift group velocity is towards the $+k$ direction and therefore the spin-wave group velocity in the $+k$ direction is faster than in the $-k$ direction. Consequently, $\delta v_{\text{g}}$ shows a positive sign and is twice the $v_{\text{g}}^{\text{DMI}}$. Based on Eq.~\ref{chiral} and considering an $M_{\text{S}}=141~$kA/m~\cite{Liu2014}, we can deduce a DMI constant $D=16~\mu\text{J/m$^{2}$}$. The origin of a finite $H_{0}$ is however unclear. It is not observed in the micromagnetic simulations~\cite{oommf} (green dashed line in Fig.~\ref{fig3}) showing a sharp step-like field dependence~\cite{SI}. Considering that the magnetization loop measured by the vibrating sample magnetometer~\cite{SI} is not fully closed at low field, we speculate $H_{0}$ to be a finite field required to completely eliminate small domain structures induced by surface roughness.\\
\\
\indent To study the $k$ dependence, we integrated coplanar waveguides (CPWs) on the 7 nm-thick YIG film~\cite{SI}. Two distinct modes are observed and attributed to the $k_1=3.1$ rad/$\mu$m and $k_2=9.1$ rad/$\mu$m, identified by Fourier transformation~\cite{Vla2008,Neu2010,Yu2014,Lee2016,SI}. We summarize data from NSL and CPW samples in Fig.~\ref{fig4}(a) and observe clear asymmetry. The spin-wave dispersion relation $f(k)$ [Fig.~\ref{fig4}(a)] is calculated based on
\begin{widetext}
\begin{equation}\label{dispersion}
f=\frac{\gamma\mu_{0}}{2\pi}\left[\left(H+\frac{2A}{\mu_{0}M_{\text{S}}}k^2\right)\left(H+M_{\text{S}}+\frac{2A}{\mu_{0}M_{\text{S}}}k^2\right)+\frac{M_{\text{S}}^2}{4}\left(1-e^{-2kt}\right)\right]^{\frac{1}{2}}+\left[\left(\hat{\textbf{n}}\times\hat{\textbf{H}}\right)\cdot\hat{\textbf{k}}\right]\frac{\gamma D}{\pi M_{\rm{S}}}k,
\end{equation}
\end{widetext}
where $\gamma$ the gyromagnetic ratio, $A=0.37\times10^{-11}$~J/m the exchange stiffness constant~\cite{Stancil}, $t=7$ nm the film thickness. The first term is the non-chiral contribution from the dipole-exchange spin waves~\cite{Kal1986} in a DE configuration~\cite{DE1961,Stancil,Bay2005}. The second term originates from the interfacial DMI. The chirality is determined by the vector relation $(\hat{\textbf{n}}\times\hat{\textbf{H}})\cdot\hat{\textbf{k}}$ and its amplitude is decided by the DMI strength $D$. By taking the DMI constant $D=16~\mu$J/m$^2$ extracted from $\delta v_{\text{g}}$, the calculated dispersion relations agree reasonably well with the experimental data points [Fig.~\ref{fig4}(a)]. We rotate $H$ in the film plane with respect to $k$ and measure $\delta v_{\text{g}}$ as a function of $\theta$ [Fig.~\ref{fig4}(b)]. The sinusoidal angular dependence is observed consistent with the vector relation $(\hat{\textbf{n}}\times\hat{\textbf{H}})\cdot\hat{\textbf{k}}$. In addition to the 7 nm-thick YIG film, we also measured spin-wave propagation in YIG films with other thicknesses~\cite{SI}. The thickness dependence of $\delta v_{\text{g}}$ is shown in Fig.~\ref{fig4}(c), where the asymmetry in group velocities enlarges with decreasing thickness. The observed thickness dependence is in opposite to that of the surface anisotropy-induced nonreciprocity~\cite{Gla2016,Oli2019,SI}. This observation indicates that the observed DMI is an interfacial effect. The DMI constant $D$ can then be expressed as $D=\frac{\lambda D_{\text{i}}}{t}$~\cite{Lee2016}, where $D_{\text{i}}$ the interfacial DMI parameter and $\lambda$ the characteristic length of the interface, which depends on the details of the interface, such as roughness. With a rough estimation of $\lambda$ being the lattice constant of YIG $a=12.4~\text{\AA}$~\cite{Wu_book} and a linear fitting of Fig.~\ref{fig4}(c), we derive an interfacial DMI parameter $D_{\text{i}}=90~\mu$J/m$^2$. The DMI-induced frequency shifts $\delta f$~\cite{Nem2015,Lee2016,Li2017} are also observed~\cite{SI}.
\\
\indent The transmission spectra shown in Fig.~\ref{fig2} exhibit a clear amplitude nonreciprocity. The nonreciprocity parameter $\beta=\frac{S_{12}-S_{21}}{S_{12}+S_{21}}$~\cite{TYu2019} is extracted as a function of $\theta$ (Fig.~\ref{fig4}(d)). At $\theta=90^\circ$, $\beta$ increases when films go thicker~\cite{SI} due to the DE nonreciprocity~\cite{Wong2014}. The remaining sizable amplitude nonreciprocity for 7 nm-thick YIG film may also be resulted from the asymmetric decay of propagating spin waves~\cite{Kam1975}, but most likely due to the excitation characteristics given by the antenna~\cite{Demi2009,Seki2010}. Interestingly, an unexpected increase of nonrecirpocity occurs around 30$^\circ$ [Fig.~\ref{fig4}(d)]~\cite{SI}, which may result from the interfacial DMI. The spin-wave dispersions at 30$^\circ$ become rather flat and therefore only a small group velocity $v_{\text{g}}^{0}$ remains~\cite{Kal1986,Bay2005,Qin2018,SI}. If the DMI-induced drift group velocity $v_{\text{g}}^{\text{DMI}}\approx v_{\text{g}}^{0}$, the nonreciprocity is enhanced due to the interfacial DMI~\cite{Pirro2017,SI}.\\ 
\indent
We confirmed the interfacial DMI in ultrathin YIG films independently by BLS. The BLS spectra measured on the 10-nm thick YIG showed clear frequency shifts $\delta f$ between the Stokes and Anti-Stokes peaks~\cite{SI}, indicating an asymmetric dispersion $f(k)$ induced by the DMI. We extract $\delta f$ from the BLS data and plot them as a function of the spin-wave wavevector $k$ in Fig.~\ref{fig5}, where $\delta f$ increases with an increasing $k$. This is consistent with the linear relationship given by $\delta f=\frac{2\gamma}{\pi M_{\text{S}}}Dk$~\cite{Cor2013,Moon2013,Nem2015,Lee2016,Li2017}. Fitting the $k$ dependent frequency shifts we extract a DMI constant of 10.3$\pm 0.8$~$\mu$J/m$^{2}$ for the 10 nm-thick sample. This value is in good agreement with 9.9$\pm 1.9$~$\mu$J/m$^{2}$ extracted from the chiral spin-wave velocities measured by the AESWS [Fig.~\ref{fig4}(c)]. The BLS spectra taken on a 7 nm-thick YIG film exhibit a small signal-to-noise ratio~\cite{SI}. Still we observe a clear frequency shift of up to about 0.15 GHz at $k=13.3$~rad/$\mu$m that yields a DMI constant of 14.2$\pm 4.2$~$\mu$J/m$^{2}$.\\

\indent We now discuss the origin of the interfacial DMI in ultrathin YIG films. In general, the interfacial DMI stems from the inversion symmetry breaking in the film normal direction $\hat{\textbf{n}}$ and the spin-orbit coupling. In these samples, the upper surface is either exposed to air or covered by antennas. The observed effect does not change with different adhesion layers~\cite{SI}, which indicates that the DMI does not originate from the top but from the bottom surface of YIG consistent with TmIG/GGG samples reported recently~\cite{Beach2019,Ding2019,Velez2019}. The DMI may be enhanced by a heavy metal layer~\cite{Ma2018,Shao2019} on YIG, but consequently the damping will be severely affected~\cite{Sun2013}. The energy dispersive X-ray spectroscopy and geometric phase analysis~\cite{SI} are conducted and no significant Gd diffusion~\cite{Gom2018} is found in the 7 nm-thick YIG film grown by sputtering. Oxide interfaces are demonstrated experimentally to exhibit Rashba splitting, for example in LaAlO$_3$/SrTiO$_3$~\cite{Cav2010}. The Rashba-induced DMI is predicted at oxide-oxide interfaces~\cite{Ban2013} and is further demonstrated by simulations to form chiral magnetic textures, such as skyrmions~\cite{Ban2014}. It has been manifested by first-principles calculations that Rashba-induced DMI exists in graphene-ferromagnet interface~\cite{Yang2018}, even without strong spin-orbit coupling. For semiconductor heterostructures consisting of materials with different band gaps it is reported that the band offsets in conductance (and valence bands) are relevant for the Rashba effect~\cite{Dirk2000}. The insulators YIG and GGG exhibit different band gaps~\cite{Onb2016} and corresponding band offsets are likely. To fully understand the origin of the interfacial DMI at the YIG/GGG interface, more studies such as first-principals calculations~\cite{Yang2018,Mertig2019} and X-ray magnetic dichroism~\cite{PYu2010,Ingrid2018} are required, which are beyond the scope of this work. It would also be instructive to study the substrate dependence of the spin-wave nonreciprocity. However, it proves to be highly challenging to fabricate low-damping YIG on conventional substrates, such as Si~\cite{Che2016} and GaAs~\cite{Sad2019}. The nonreciprocity effect induced by the dipolar interaction between top and bottom layers should be negligible since the parallel magnetic configuration can be established with a small field~\cite{SI,Gal2019}\\ 
\indent 
To summarize, we have observed chiral spin-wave velocities in ultrathin YIG films induced by DMI attributed to the YIG/GGG interface. The drift group velocity is about 40~m/s yielding a DMI constant of 16~$\mu$J/m$^2$. The chirality is ruled by the vector relation $(\hat{\textbf{n}}\times\hat{\textbf{H}})\cdot\hat{\textbf{k}}$, verified by angle-dependent measurements. The DMI-induced chiral propagation of spin waves in the magnetic insulator YIG offers great prospects for chiral and spin-texture based magnonics~\cite{Schwarze2015,Xiao2015,Wagner2016,Hama2018}.\\

\begin{acknowledgments} 
The authors thank R. Duine, M. Kuepferling and A. Slavin for their helpful discussions and H.-Z. Wang for her help on the illustration. Financial support by NSF China under Grant Nos. 11674020 and U1801661, 111 talent program B16001, the National Key Research and Development Program of China No. 2016YFA0300802 and 2017YFA0206200, ANR-12-ASTR-0023 Trinidad and by SNF via project 163016 and sinergia grant 171003 Nanoskyrmionics is gratefully acknowledged. T.L. and M.W. were supported by the U.S. National Science Foundation (EFMA-1641989) and the U.S. Department of Energy, Office of Science, Basic Energy Sciences (DE-SC0018994). Y.L. and P.G. were supported by National Natural Science Foundation of China (51672007, 11974023), and The Key R$\&$D Program of Guangdong Province (2018B030327001, 2018B010109009).
\end{acknowledgments}
\providecommand{\noopsort}[1]{}\providecommand{\singleletter}[1]{#1}%

\end{document}